\documentstyle[12pt,aasms4]{article}
\tightenlines

\def\teff {$T_{\rm eff}$}

\begin{document}

\title{The Extent and Cause of the Pre-White Dwarf Instability Strip}

\author{ M.~S.~O'Brien\altaffilmark{1} }

\altaffiltext{1}{Space Telescope Science Institute, 
3700 San Martin Drive, Baltimore, MD 21218; obrien@stsci.edu.}

\begin{abstract}

One of the least understood aspects of white dwarf evolution is the
process by which they are formed.  The initial stages of white dwarf
evolution are characterized by high luminosity, high effective temperature,
and high surface gravity, making it difficult to constrain their properties 
through traditional spectroscopic observations.  We are aided, however, 
by the fact that many H- and He-deficient pre-white dwarfs (PWDs) are 
multiperiodic $g$-mode pulsators.  These stars fall into two classes, 
the variable planetary nebula nuclei (PNNV) and the ``naked'' GW~Vir stars.
Pulsations in PWDs provide a unique opportunity to probe their interiors, 
which are otherwise inaccesible to direct observation.  Until now,
however, the nature of the pulsation mechanism, the precise 
boundaries of the instability strip, and the mass distribution of the
PWDs were complete mysteries.  These problems must be addressed before
we can apply knowledge of pulsating PWDs to improve understanding of 
white dwarf formation.

This paper lays the groundwork for future theoretical investigations of
these stars.   In recent years, Whole Earth Telescope observations led to
determination of mass and luminosity for the majority of the GW~Vir 
pulsators.  With these observations, we identify the common properties 
and trends PWDs exhibit as a class.  

We find that pulsators of low mass have higher luminosity, suggesting 
the range of instability is highly mass-dependent.  The observed trend 
of decreasing periods with decreasing luminosity matches a decrease in 
the maximum (standing-wave) $g$-mode period across the instability 
strip.  We show that the red edge can be caused by the lengthening 
of the driving timescale beyond the maximum sustainable period.  This 
result is general for ionization-based driving mechanisms, and it 
explains the mass-dependence of the red edge.  The observed form of the 
mass-dependence provides a vital starting point for future theoretical 
investigations of the driving mechanism.  We also show that the blue 
edge probably remains undetected because of selection effects arising
from rapid evolution.  

\end{abstract}


\section{Introduction}

Until about twenty years ago, the placement of stars in the transition
region between the planetary nebula nucleus track and the upper end of 
the white dwarf cooling sequence was problematic.  This was due not only 
to the rapidity with which stars must make this transition (making 
observational examples hard to come by) but also to the difficulty of 
specifying log~$g$ and $T_{\rm eff}$ for such objects.  Determining these 
quantities from spectra requires that we construct a reasonable model of the 
star's atmosphere.  This is very difficult for compact stars with 
$T_{\rm eff}$ in excess of 50,000~K.  The assumption of local thermal 
equilibrium (LTE), so useful in modeling the spectra of cooler stars, breaks 
down severely at such high temperatures and gravities.

Fortunately, the known sample of stars that occupy this phase of
the evolutionary picture has grown over the last
two decades.  The most important discovery was that of a new
spectral class called the PG~1159 stars.  They are defined by the 
appearance of lines of HeII, CIV, and OVI (and sometimes NV) in their 
spectra.  Over two dozen are known, ranging in $T_{\rm eff}$ from
over 170,000~K down to 80,000~K.  About half are central stars of
planetary nebula.  The most evolved PG~1159 stars merge with the
log~$g$ and $T_{\rm eff}$ of the hottest normal white dwarfs.  This
class thus forms a complete evolutionary sequence from PNN to the
white dwarf cooling track (Werner 1995, Dreizler \& Huber 1998).

About half of the PG~1159 stars are pulsating variable stars, spread
over the entire range of log~$g$ and $T_{\rm eff}$ occupied by members of the
spectral class.  This represents the widest instability ``strip''
(temperature-wise) in the H-R diagram.  Central star variables are
usually denoted as PNNV stars (planetary nebula nucleus variables).
Variable PG~1159 stars with no nebula make up the GW~Virginis (or
simply GW~Vir) stars.  PG~1159 thus serves as the prototype for both a
spectroscopic class and a class of variable stars.  Farther down the
white dwarf sequence, we find two additional instability strips.  At surface
temperatures between about 22,000 and 28,000~K (Beauchamp et 
al.~1999), we find the DBV (variable
DB) stars.  Even cooler are the ZZ~Ceti (variable DA) stars, with
$T_{\rm eff}$ between 11,300 and 12,500~K (Bergeron et al.~1995).
Variability in
all three strips results from $g$-mode pulsation (for the ZZ~Cetis, see
Warner \& Robinson 1972; for the DBVs, see Winget et
al.~1982; for the PG~1159 variables, see Starrfield et al.~1983).  The
pulsation periods provide a rich mine for probing the structure
of white dwarf and pre-white dwarf (PWD) stars.

Despite the wealth of pulsational data available to us in studying
the variable PWDs, they have so far resisted coherent
generalizations of their group properties. Such a classification is
required for understanding possible driving mechanisms or explaining
the observed period distribution.  Until now, the errors bars
associated with spectroscopic determination of mass, luminosity, and \teff~for a
given variable star spanned a significant fraction of the
instability strip.

Even given the limited information provided by spectroscopic
determinations of log~$g$ and \teff, many attempts have been made
to form a coherent theory of their pulsations.  Starrfield et al.~(1984)
proposed that GW~Vir and PNNV stars are driven by cyclical ionization 
of carbon (C) and oxygen (O).  However, their proposal suffers from the 
difficiency that a helium (He) mass fraction of only 10\% in the driving 
zone will ``poison'' C/O driving.  Later, they 
managed to create models unstable to (C driven) pulsation with surface He 
abundance as high as 50\%, but only at much lower $T_{\rm eff}$ than most 
GW~Vir stars (Stanghellini, Cox, \& Starrfield~1991; see also the review
by Cox, 1993).
Another problem is the existence of non-pulsating stars within 
the strip with nearly identical spectra to the pulsators (Werner 1995).  
More precise observations can detect subtle differences between the 
pulsators and non-pulsators.  For instance, Dreizler (1998) finds
NV in the spectra of all the pulsators but only some of the 
non-pulsators.  It remains to be seen if these differences are important.

More recently, Bradley \& Dziembowski (1996) studied the effects of
using the newer OPAL opacities in creating unstable stellar models.
Their models pulsate via O driving, and---in exact opposition to 
Starrfield, Stanghellini, and Cox---they require {\it no} C (or He)
in the driving region to obtain unstable modes that match the observed
periods.  Their models also require radii up to 30\% larger than those
derived from prior asteroseismological analyses of GW~Vir stars, in
order to match the observed range of periods.

Finally, Saio (1996) and Gautschy (1997) proposed driving by a 
``metallicity bump,'' where the opacity derivative 
changes sign due to the presence of metals such as iron in the envelope.  This 
$\kappa$~mechanism is similar to those currently thought to drive pulsation 
in the $\beta$-Cephei variables (see for instance Moskalik \& 
Dziembowski, 1992) and in the recently discovered subdwarf~B 
variables (Charpinet et al.~1996).  Unfortunately, their
simplified models are inconsistent with the evolutionary status of
PG~1159 stars.  More importantly, their period structures do not 
match published WET observations of real pre-white dwarfs (Winget et al.~1991,
Kawaler et al.~1995, Bond et al.~1996, O'Brien et al.~1998).

With so many different explanations for pulsational driving, stricter 
constraints on the observable conditions of the pulsators and 
non-pulsators are badly needed.  The most effective way to thin the 
ranks of competing theories (and perhaps point the way to explanations 
previously unthought of) is to obtain better knowledge of when the 
pulsations begin and end for PWD stars of a given mass.

Of course, with only a few stars available to study, even complete 
asteroseismological solutions for all of them might not prove
significantly illuminating.  Even if their properties follow 
recognizable patterns, it is difficult to show this compellingly 
given only a few cases.  However, with asteroseismological analyses 
now published for the majority of the GW~Vir stars, we can finally 
attempt to investigate their behavior as a class of related objects.  
In the next section, we outline the analytic theory of PWD pulsation.
In \S~3, we use the observed properties of the variable PWDs to 
show that they do follow compelling trends, 
spanning their entire range of stellar parameters, to which any model of PWD 
pulsation must conform.  Next we introduce a new set of 
numerical models developed to help interpret this behavior.  
In \S~5, we show how the evolution (and 
eventual cessation) of pulsation in PWDs can be governed by the 
changing relationship of the driving timescale to the maximum 
sustainable $g$-mode period.  These results suggest several fruitful 
directions for future work, both theoretical and observational, which 
we discuss in the concluding section.

\section{Theory}

The set of periods excited to detectable limits in white dwarf stars
is determined by the interplay of several processes.  The first is the
driving of pulsation by some---as yet unspecified---mechanism (no
matter how melodious the bell, it must be struck to be heard).  Second
is the response of the star to the driving taking place somehere
in its interior.  A pulsating PWD star is essentially a
(spherically symmetric) resonant cavity, capable of sustained vibration
at a characteristic set of frequencies.  Those frequencies are
determined by the structure of the star, its mass and luminosity,
as well as the thickness of its surface layers.  Finally, the actual
periods we see are affected by the mechanism through which internal
motions are translated into observable luminosity variations.   This
is the so-called ``transfer function,'' and clues to its nature are
to be found in the observed variations as well.

\subsection{Asyptotic Relations}

If we wish to make the most of the observed periods, we must understand
the process of driving, and response to driving, in as much detail as 
possible.  However, we can
learn a great deal by simply comparing the periods of observed light variations
to the normal-mode periods of model stars.  

The normal mode oscillations of white dwarf and PWD stars are
most compactly described using the basis set of spherical harmonic functions,
$Y^{\ell}_{m}$, coupled with an appropriate set of radial wave functions
$R_{n}$.  Here $n$ is the number of nodal surfaces in the radial direction,
$\ell$ denotes the number of nodal planes orthogonal to such surfaces,
and $m$ is the number of these planes which include the pulsational axis
of the star.

The periods of $g$-modes of a given $\ell$ are expected to increase
monotonically as the number of radial nodes $n$ increases.  The reason
is that the buoyant restoring force is proportional to the total mass
displaced, and this mass gets smaller as the number of radial nodes
increases.  A weaker restoring force implies a longer period
(see, e.g., Cox 1980).  In the ``asymptotic limit'' that $n\gg\ell$, the periods of
consecutive modes should obey the approximate relation
\begin{equation}
\Pi_{n} \cong \frac{\Pi_{\rm 0}}{[\ell(\ell+1)]^{1/2}} (n +
\epsilon)~~~~~~~~~~~~~~~n\gg\ell,
\end{equation}
where $\Pi_{n}$ is the $g$-mode period for a given value of $n$, and
$\Pi_{\rm 0}$ is a constant that depends on the overall structure of
the star (see, e.g., Tassoul 1980, Kawaler 1986).\footnote
{The additional constant $\epsilon$ is assumed to be small,
though its exact value depends on the boundary conditions.
Since the actual boundary conditions depend on the period, $\epsilon$
probably does, too.}

Equation (1) implies that
modes of a given $\ell$ should form a sequence based upon a
fundamental period spacing $\Delta \Pi = \Pi_{\rm 0}/\sqrt{\ell(\ell+1)}$.
This is the overall pattern identified by various investigators in the 
lightcurves of most of the GW~Vir stars.  Once a period spacing 
is identified, we can compare this spacing to those computed in models to decipher the star's
structure.  Kawaler (1986) found that the
parameter $\Pi_{\rm 0}$ in static PWD models is dependent
primarily on the overall stellar mass, with a weak
dependence on luminosity. Kawaler \& Bradley (1994) present the approximate
relation
\begin{equation}
\Pi_0 \cong 15.5 \left(\frac{M}{M_{\odot}}\right)^{-1.3}
\left(\frac{L}{100L_{\odot}}\right)^{-0.035}
\left(\frac{q_{y}}{10^{-3}}\right)^{-0.00012}
\end{equation}
where $q_{y}$ is the fraction by mass of He at the surface.\footnote
{Note that the sign of the exponent in the $L$ term is in error
in Kawaler \& Bradley (1994).}

Other questions, however, can only be answered from knowledge of the cause of the
light variations we measure.  In the case of the PWDs, this
is chief among the mysteries we would like to solve.  The most telling
clue will be the extent and location of the region of instability in the
H-R diagram---derived in part from pulsational analysis of the structure
of stars which bracket this region. The will now provide some of
the background needed to attack these issues.

\subsection{Pulsation in Pre-White Dwarf Stars}

In a star, energy generally flows down the temperature
gradient from the central regions to the surface in a smooth,
relatively unimpeded fashion.  Of course small, random perturbations to this
smooth flow constantly arise.  The situation is stable as long as
such perturbations quickly damp out, restoring equilibrium.  For instance, 
equilibrium is restored by the forces of buoyancy and pressure; theses forces 
define the nature of $g$- and $p$-modes.  They resist mass motions and local 
compression or expansion of material away from equilibrium conditions.

\subsubsection{Driving and Damping}

Thermodynamics and opacity affect local
equilibrium.  In general, if a parcel of material in a star is compressed, its
temperature goes up while its opacity decreases.  The higher temperature
causes more radiation to flow out to the surrounding material, while
lower opacity decreases the efficiency with which radiation is absorbed.
From the first law of thermodynamics, an increasing temperature
accompanied by net heat loss implies that work is being done on the parcel
by its surroundings.  Similar arguments show that work is done on the
parcel during expansion, also.  Thus any initial perturbation will be
quickly damped out---each parcel demands work to do otherwise.  The
requirement that a region lose heat when compressing and gain
heat when expanding is the fundamental criterion for stability.  When
the opposite is true, and work is done by mass elements on their
surroundings during compression and expansion, microscopic perturbations can
grow to become the observed variations in pulsating stars.

Under certain circumstances, the sign of the opacity derivative changes
compared to that described above.  If the opacity, $\kappa$, increases
upon compression, then heat flowing through a mass element is trapped
there more efficiently.  Within regions where this is true, work is
done on the surrounding material.  Thus, these regions can help destabilize
the star, if this driving is not overcome elsewhere in the star.  However,
other regions, where work is required to compress and expand material,
tend to damp such pulsation out.  Global instability arises only
when the work performed by the driving regions outweighs the
work done on the damping regions over a pulsation cycle.  In this
case, the flow of thermal energy can do mechanical work, and this work
is converted into the pulsations we observe.

This method of driving pulsation is called the $\kappa$ mechanism.
A region within a star will drive pulsation via this mechanism
if the opacity derivatives satisfy the condition (see for instance
Cox 1980)
\begin{equation}
\frac{d}{dr}\left(\kappa_{T}+\frac{\kappa_{\rho}}{\Gamma_{3}-1}\right)>0
\end{equation}
where
\[
\kappa_{T} \equiv \left(\frac{\partial ln \kappa}{\partial ln T}\right)_{\rho},~~~~~~~
\kappa_{\rho} \equiv \left(\frac{\partial ln \kappa}{\partial ln \rho}\right)_{T},~~~{\rm and}~~~~
\Gamma_{3}-1 \equiv \left(\frac{\partial ln T}{\partial ln \rho}\right)_{S}.
\]
Here $S$ represents the specific entropy, $\kappa$ is the opacity in cm$^{2}$/g, and
the variables $r$, $\rho$, and $T$ all have their usual meaning.

Equation~(3) is satisfied most commonly when some species within a
star is partially ionized.  In particular, $\kappa_{T}$ usually increases
in the hotter (inner) portion of a partial ionization zone and decreases
in the cooler (outer) portion.  Thus the inner part of an ionization
zone may drive while the outer part damps pulsation.  The
adiabatic exponent, $\Gamma_{3}-1$, is always positive but usually
reaches a minimum when material is partially ionized.  This enhancement of the
$\kappa$ mechanism is called the $\gamma$ mechanism.  Physically,
the $\gamma$ mechanism represents the conversion of some of the work
of compression into further ionization of the species in question.
This tends to compress the parcel more, aiding the instability.
Release of this ionization energy during expansion likewise increases
the purturbation.  Since they usually occur together, instabilities
caused by both the opacity and ionization effects are known as the
$\kappa$-$\gamma$ mechanism.

The pulsations of Cepheid and RR~Lyrae variables, for instance, are
driven via the $\kappa$-$\gamma$ mechanism operating within a region
where HeI and HeII have similar abundances. This same partial ionization
zone is apparently the source of instability for the DBV white dwarfs.
The variations observed in ZZ~Ceti white dwarfs have long been attributed to
partial ionization of H.  However, Goldreich \& Wu (1999a,b,c) have
recently shown the ZZ~Ceti pulsations can be driven through a different
mechanism in efficient surface convection.\footnote
{Such convection zones often accompany regions of partial ionization
associated with the $\kappa$-$\gamma$ mechanism.  However, the driving
proposed by Goldreich \& Wu is not directly related to ionization.
It is possible this theory might eventually be expanded to account for
DBV pulsation as well.  It is unlikely to find application in PWDs,
though, since models of PG~1159 stars generally do not
support convection.}
Great efforts have been expended by theorists attempting to explain
GW~Vir and PNNV pulsations in terms of some combination of C and O
partial ionization (Starrfield et al.~1983; Stanghellini, Cox, \&
Starrfield~1989; Bradley \& Dziembowski~1996).  A primary difficulty 
arises from the damping effects of He (and C in the O-driving scheme
proposed by Bradley \& Dziembowski 1996) in the driving zone, which can 
``poison'' the driving.  More recently, Saio (1996) and Gautschy 
(1997) attempted to explain
driving in terms of an ``opacity bump'' in the models, without partial
ionization---in other words, using the $\kappa$ mechanism alone.  A
fundamental problem has been the lack of information on the exact extent
of the instability strip and the structure of the pulsating stars
themselves.  The initial goal in this paper, therefore, is to 
define as precisely as possible the observational attributes which any proposed
driving source must reproduce.  Whether or not the mechanism we later identify
is correct, we hope to lay the groundwork for future studies which will
eventually provide a definitive answer to this question.

If a star pulsates, we wish to know
what relationship the driving has to the periods we observe.  In
general, no star will respond to driving at just any arbitrary
period; the pulsation period range is determined by
the driving mechanism.  Cox (1980) showed that the approximate
pulsation period is determined by the thermal timescale of the
driving zone:
\begin{equation}
\Pi \sim \tau_{\rm th} = \frac{c_{\rm v}Tm_{\rm dz}}{L}.
\end{equation}
Here $\tau_{\rm th}$ is the thermal timescale, $c_{\rm v}$ is the heat
capacity, and $m_{\rm dz}$ is the mass above the driving zone.
This equation gives the approximate time it takes the star
to radiate away---via its normal luminosty, $L$---the energy contained
in the layers above the region in question.  Though Cox derived this
relationship for radial modes, Winget (1981) showed that it applies
equally well to nonradial $g$-mode pulsation.

The basic idea behind Equation~(4) is that energy must be modulated
at approximately the same rate at which it can be dammed up and released
by the driving zone.  Consider the question of whether a given
zone can drive pulsation near a certain period.  If the driving zone is
too shallow, then the thermal timescale is shorter than the pulsation
period.  Any excess heat is radiated away before the compression increases
significantly.  Thus, it can't do work
to create mechanical oscillations on the timescale in question.  If the
driving zone is too deep ($\tau_{\rm th} > \Pi$), then excess heat built
up during contraction is not radiated away quickly enough during expansion;
it then works against the next contraction cycle.  Of course, this relationship
is only approximate, and other factors might intervene to limit pulsation
to periods far from from those implied by Equation~(4).

\subsubsection{Period Limits}

One such factor is the range of pulsation modes a star can sustain in the
form of standing waves.  Hansen, Winget, \& Kawaler (1985) showed
that there is a maximum period for white dwarf and PWD
stars above which oscillations will propagate only as running waves
that quickly damp out.  They attempt to calculate this maximum $g$-mode
period, $\Pi_{\rm max}$, to explain the observed trend of
ZZ~Ceti periods with $T_{\rm eff}$.  We can recast their equations 
(6) through (8) as 
\begin{equation}
\Pi_{\rm max} \approx  940s~\left(\frac{\mu}{\ell(\ell+1)}\right)^{0.5} 
\left(\frac{R}{0.02R_{\odot}}\right) 
\left(\frac{T_{\rm eff}}{10^{5}{\rm K}}\right)^{-0.5},
\end{equation}
or, using the relation $R^{2} = L / 4\pi \sigma T_{\rm eff}^{4}$,
\begin{equation}
\Pi_{\rm max} \approx 940s~\left(\frac{\mu}{\ell(\ell+1)}\right)^{0.5} 
\left(\frac{L}{35L_{\odot}}\right)^{0.5} 
\left(\frac{T_{\rm eff}}{10^{5}{\rm K}}\right)^{-2.5}.
\end{equation}
where $R$ is the stellar radius, $\mu$ represents the mean molecular weight, 
and $\ell$ is the pulsation index introduced earlier.   For a complete 
derivation of these equations, see the Appendix.

For cool white dwarfs of a given mass, the radius is roughly
constant with time, and $\Pi_{\rm max}$ is expected to increase as stars
evolve to lower $T_{\rm eff}$ and the
driving zone sinks deeper.  Cooler ZZ~Ceti and DBV white dwarf stars should in general
have longer periods; for the ZZ~Cetis at least, this is indeed the case.

However, in PWDs degeneracy is
not yet sufficient to have halted contraction, and the $R$ dependence
is still a factor in determining how $\Pi_{\rm max}$ varies
through the instability strip.  We cannot say {\it a priori} which dominates: the
shrinking radius which tends to decrease $\Pi_{\rm max}$, or the
falling $T_{\rm eff}$ which has the opposite effect.  We cannot even
predict in advance whether $\Pi_{\rm max}$ is a factor in determining the
pulsation periods at all. In \S~5, we answer these questions in
view of the observed properties of both the PNNV and GW~Vir stars.

In summary, we can measure individual white dwarf mass
and luminosity through identification of the period spacing. 
The resulting determinations of GW~Vir structure, summarized in the next 
section, will tell us the precise boundaries
of the instability strip, in other words, when the pulsations begin and end
in the evolution of PWD stars of a given mass.  This knowledge, coupled with the
timescales of driving and the maximum $g$-mode period in models matched to the
observations, will provide strict constraints on the allowable
form of the driving mechanism.  This information is absolutely necessary to
any research program designed to discover why PWDs pulsate in the first place.

\section{Observed Temperature Trends of the PG~1159 Pulsators}

\subsection{Mass Distribution}

White dwarfs exhibit a very narrow mass distribution, centered at about 
0.56--0.59$~M_{\odot}$ (Bergeron, Saffer \& Liebert 1992, Weidemann
\& Koester 1984).  If, for a given \teff, pulsating PWD 
masses conform to the same distribution expected of non-variables, then 
we are led to certain expectations concerning the pulsations seen in 
the former.  Based on Equations~(1) and (2), period spacing (proportional to 
$\Pi_{\rm 0}$) should increase monotonically as the luminosity and \teff~of
a given star decrease.  

Four PWD stars have so far yielded to
asteroseismological scrutiny, the GW~Vir stars PG~1159 (Winget et al.~1991), PG~2131
(Kawaler et al.~1995), and PG~0122 (O'Brien et al.~1998), and the central star of 
the planetary nebula NGC~1501 (Bond et al.~1996).
We summarize the parameters of these four stars in Table~1.
All show patterns of equal period spacing very close to $21~$s.
This is a remarkable trend, or more accurately, a remarkable lack of a trend!
If these stars follow the narrow mass distribution observed in white
dwarf stars, then the period spacing is expected to increase with decreasing
\teff~as they evolve from the blue to the red edge of the instability
strip.  Observationally, this is not the case.

For instance PG~1159, with a \teff~of 140,000~K, 
should see its period spacing increase by about 20\%,
from 21~s to 26~s, by the time it reaches the effective
temperature of PG~0122---80,000~K.  In other words, the
farther PG~0122's period spacing is from 26~s, the farther
is its mass from that of PG~1159.  In fact these two
objects, representing the high and low \teff~extremes of the 
GW~Vir stars, have almost exactly the {\it same} period spacing 
despite enormous differences in luminosity and temperature.  
For PG~0122, its low \teff~pushes it toward longer $\Delta\Pi$;
this must be offset by a higher mass.  With such a significant 
\teff~difference, the mass difference between PG~0122 and PG~1159 must be 
significant also, and it is: $0.69~M_{\odot}$ versus $0.58~M_{\odot}$.

Two stars with a coincident period spacing---despite widely
different mass and luminosity---might simply be curious.  In fact 
all four PWDs with relatively certain
period spacing determinations have the same spacing to within
2~s, or 10\%.  This includes the central star of NGC~1501, which
has a luminosity over three orders of magnitude larger than that 
of PG~0122. Hence, NGC~1501 must have an even lower mass than PG~1159 
by comparison.  Figure~1 shows the mass versus luminosity values for 
the known GW~Vir stars plus NGC~1501, based on the values 
of $\Delta\Pi$ from Table~1. 

Figure~2 shows the implications of the common 21-22~s
spacing for the instability ``strip'' in the 
log~$g$--log~\teff~plane.  The observational region
of instability has shrunk significantly.  It
exhibits such a striking slope in the figure that, unlike most other
instability regions in the H-R diagram, it can no longer be 
refered to accurately as an instability strip (in 
temperature) at all.  Nevertheless, we will continue to refer
to the region of instability pictured in Figure~2 as
the GW~Vir ``instability strip'' with the understanding
that the effective temperatures of the red and blue edges are 
highly dependent on log~$g$ (or $L$).

Why should the PWD instability strip apparently
straddle a line of approximately constant period spacing?
Normally, theorists search for explanations for the observed
boundaries (the red and blue edges) of an instability strip
based on the behavior of a proposed pulsation mechanism. That behavior
is determined by the thermal properties of a PWD
star, while its period spacing is determined by its mechanical
structure.  In degenerate or nearly degenerate stars, the
thermal properties are determined by the ions, while the
mechanical structure is determined by the degenerate
electrons, and normally the two are safely treated as
separate, isolated systems.  If the 21-22~s period spacing
is somehow a {\it prerequisite} for pulsation, then
this implies an intimate connection between the mechanical
oscillations and the thermal pulsation mechanism.

The alternative is that the mass-luminosity relationship along
the instability strip is caused by some other process---or 
combination of processes---which approximately coincides with the
relationship that governs the period spacing.  In
this case, some mechanism must shut off observable pulsation
in low mass PWDs before they reach low temperature, 
and delay observable pulsation in higher mass PWDs 
until they reach low temperature.\footnote
{We use the phrase ``observable pulsation'' to indicate that 
possible solutions might reside in some combination of observational 
selection effects as well the intrinsic behavior of a driving 
mechanism.}  
We will explore mechanisms which meet these criteria
in \S~5.

\subsection{Period Distribution}

The PWD pulsators exhibit another clear observational trend: their 
periods decrease with decreasing luminosity (increasing surface gravity).
Figure~3 shows the luminosity versus dominant period (that is, the period of
the largest amplitude mode) for the same stars from Figure~1.  The trend 
apparent in the figure is in marked 
contrast to the one seen in the ZZ~Ceti stars, which show longer periods 
with lower \teff.  The ZZ~Ceti period trend is generally attributed
to the changing thermal timescale in the driving zone, 
which sinks deeper (to longer timescales) as the star cools.  If the 
same effect determines the periods in GW~Vir and PNNV stars, then 
Figure~3 might be taken to indicate that the driving zone becomes 
more shallow with decreasing \teff. We will show that this is not 
the case in PWD models.  We conclude that some other 
mechanism must be responsible for setting the dominant period in PWD variables.  

Are the trends seen in Figures~1 through 3 related?  To explore 
this question in detail, we developed a new set of PWD 
evolutionary models, which we summarize in the next section.  In 
\S~5 we analyze the behavior of driving zones 
in PWD models in light of---and to seek an explanation 
for---the trends just discussed.

\section{A New Set of Pre-White Dwarf Evolutionary Models}

To understand the various trends uncovered in the hot pulsating
PWDs, we appeal to stellar models.  Models have been
essential for exploiting the seismological observation of individual
stars.  For this work, though, we needed models over the entire range
of the GW~Vir stellar parameters of mass and luminosity.  Our
principal computational tool is the stellar evolution program ISUEVO,
which is described in some detail by Dehner (1996; see also Dehner \&
Kawaler 1995).  ISUEVO is a ``standard'' stellar evolution code that
is optimized for the construction of models of PWDs and
white dwarfs.

The seed model for the models used in this section was generated with
ISUEVO by evolution of a $3~M_{\odot}$ model from the Zero Age Main
Sequence through the thermally pulsing AGB phase.  After reaching a
stable thermally pulsing stage (about 15 thermal pulses), mass loss
was invoked until the model evolved to high temperatures.  This model
(representing a PNN) had a final mass of $0.573~M_{\odot}$, and a
helium-rich outer layer.

To obtain self-consistent models within a small range of masses, we
used the $0.573~M_{\odot}$ model, and scaled the mass up or down.  For
example, to obtain a model at $0.60~M_{\odot}$, we scaled all
parameters by the factor $0.60/0.573$ for an initial model.
Relaxation to the new conditions was accomplished by taking many very
short time steps with ISUEVO.  Following this relaxation, the
evolution of the new model proceeded as before.  In this way, we
produced models that were as similar as possible, with mass being the
only difference.

Comparison of our evolutionary tracks and trends with the earlier
model grids of Dehner (1996) shows very close agreement, given the
different evolutionary histories.  Dehner's initial models were
derived from a single model with a simplified initial composition
profile, while our models are rooted in a self-consistent
evolutionary sequence.  We note that the work by Dehner (1996)
included elemental diffusion (principally by gravitational settling),
while the models we use here did not include diffusion.  Within the
temparature range of the GW~Vir stars, however, observations of their
surface abundances indicate that the effects of diffusion have only a 
small influence.

\section{Selection Effects, Driving, and the Blue and Red Edges}

\subsection{The Observed ``Blue'' and ``Red'' Edges}

In explaining the observed distribution of pulsating stars with respect
to stellar parameters, we must distinguish observational selection effects 
from causes intrinsic to the objects under study.  Usually, understanding
selection effects is an important part of decoding the shape of the distribution
in terms of physical effects.  In this case, the blue and red edges
exhibit a similar slope in the log~$g$--log~\teff~plane, but we must still 
separate out selection effects from the intrinsic shape of one or both of them.

The more rapid the evolution through a 
particular region of the H-R diagram, the less likely it is that stars will be 
found there.  Also, the relative sample volume is larger for stars with higher 
luminosity, since they are detectable at greater distances.  
Some combination of these effects will 
determine the odds of finding stars of a particular mass at a particular point 
in their evolution.  One of the most common ways to explore these combined 
effects is to construct a luminosity function, which is simply a plot of the 
expected number density of stars per unit luminosity, based on how bright
they are and how fast they are evolving at different times.  Figure~4 
shows schematic luminosity functions for PWD stars of two different masses, 
based on the models described in the previous section and normalized according
to the white dwarf mass distribution (see for instance Wood 1992).
One important result of this figure is that higher mass models always 
achieve a given number density per unit luminosity later (at lower $L$
and \teff) than lower mass models.  Also, the number density per unit 
luminosity increases for all models as they evolve to lower \teff, due
to the increasing amount of time spent in a luminosity bin.

These effects together imply that, {\it if} PWDs of all
mass pulsate all the way down to 80,000~K, then the distribution
of known pulsators should be skewed heavily toward those with both 
low \teff~{\it and} low mass.  We don't see such stars; thus the red edge
must exclude the low mass, low \teff~stars from the distribution.

While the observed red edge actually marks the dissappearance
of pulsations from PWD stars, selection effects
change our chances of finding high-mass, high-\teff~pulsators.  
In this case, we expect that the likehood of finding stars of a 
given mass within the instability strip will increase the closer those 
stars get to the red edge.  This would tend to render the theoretical
blue edge (as defined by the onset of pulsation in models of a given
mass) undetectible in real PWD stars---given
the small number of known variables.  In other words, stars ``bunch up''
against the red edge due to their continually slowing rate of evolution,
causing the {\it apparent} blue edge to shadow the 
slope of the red edge in the log~$g$--log~\teff~plane.  This 
could explain the approximately linear locus of pulsating stars found
within that plane implied by their tight distribution of $\Delta\Pi$.

We are left to explain the observed red edge in terms of the intrinsic
properties of the stars themselves, which we defer to \S~5.3.
First, however, we will discuss the effects of the observed mass
distribution along the strip on the period trend seen in Figure~3.

\subsection{The $\Pi$ versus~\teff~Trend}

As mentioned previously (and as we show in \S~5.3, below), the depth of an 
ionization-based driving zone 
increases---moves to larger thermal timescales---with decreasing 
\teff~for PWD stars of a given mass.  This implies
a period trend opposite to that observed in Figure~3.  Bradley \&
Dziembowski (1996) discuss this very problem, since their models
predict that the maximum period of {\it unstable} modes should increase
as \teff~decreases.  They suggest that
the composition of the driving zone might somehow change with time, 
or that the stellar radii shrink much more quickly than is currently 
thought (or some combination of the two), in such a way as to make the 
maximum unstable period decrease
with decreasing \teff.  However, no one has yet calculated how (and
whether) these suggestions could reasonably account for the observed trend.
It is clear, however, that something other than the depth of the 
driving zone alone determines the observed period range.

What other 
mechanisms might affect the observed periods?  One such mechanism is 
the changing value of the maximum sustainable $g$-mode period, 
$\Pi_{\rm max}$.  For the ZZ~Ceti stars, $\Pi_{\rm max}$ probably does 
not influence the period distribution much, since from Equations~(5) and 
(6) it increases through the ZZ~Ceti instability strip. 
In PWDs, however, the $R$ dependence in 
Equation~(5) must be taken into account; we cannot be certain 
that the trend implied by a lengthening driving timescale won't 
find itself at odds with a decreasing $\Pi_{\rm max}$.

Figure~5 shows the (arbitrarily normalized) run of $\Pi_{\rm max}$
versus \teff~for three PWD model sequences of different mass.
Clearly, $\Pi_{\rm max}$ decreases significantly as models evolve along
all three sequences, with high-mass stars exhibiting a smaller
$\Pi_{\rm max}$ than low-mass stars at all \teff.\footnote
{If this trend continued all
the way to the cooler white dwarf instability strips, then ZZ~Ceti stars
could only pulsate at very short periods, but \teff~begins to dominate 
below around 60,000 to 70,000~K, pushing $\Pi_{\rm max}$ back to longer
and longer periods once the stars approach their minimum radius
at the top of the white dwarf cooling track.}
These two effects, when combined with the PWD
mass distribution, imply that $\Pi_{\rm max}$ should plummet precipitously
with increasing log~$g$ in the models shown in Figure~2.  
For example, the ratio of $\Pi_{\rm max}$ for PG~1159 to that 
of PG~0122 is expected to be $\sim 1.42$:1, while the ratio of
their observed dominant periods is 540:400 = 1.35:1.   The period
distribution seen in Figure~2 is thus consistent with the idea that
the value of $\Pi_{\rm max}$ determines the dominant
period in GW~Vir stars.  As we will see in the next section,
$\Pi_{\rm max}$ probably also plays an important part in
determining the more fundamental question of {\it when} a given star 
is likely to pulsate.

\subsection{Driving Zone Depth, $\Pi_{\rm max}$, and the Red Edge}

In \S~2.2, we discussed the relationship between the depth of
the driving zone and the period of $g$-mode oscillations.
Equation~(4) implies that the dominant period should increase
in response to the deepening driving zone, as long
as other amplitude limiting effects do not intervene.  Figure~5 
shows how one particular effect---the decreasing maximum period---might
reverse the trend connected to driving zone depth, and the observed
periods of GW~Vir stars supports the suggestion that $\Pi_{\rm max}$
is the key factor in setting the dominant period.  While $\Pi_{\rm max}$
limits the range of periods that can respond to driving, $\tau_{\rm th}$
in the driving zone limits the periods that can be driven.  However,
$\tau_{\rm th}$ increases steadily for all GW~Vir stars, and
$\Pi_{\rm max}$ decreases steadily.  Eventually, therefore, every 
pulsator will reach a state where $\tau_{\rm th} > \Pi_{\rm max}$
over the entire extent of the driving zone.  In such a state, the
star can no longer respond to driving at all, and pulsation will cease.
If $\Pi_{\rm max}$ remains the most important amplitude limiting
factor for stars approaching the red edge, then the red edge itself
could be caused by the situation just described.

We can test this idea by asking if it leads to the kind of 
mass-dependent red edge we see.  To answer this question, we need to 
know how the depth of the driving zone (as measured 
by $\tau_{\rm th}$) 
changes with respect to $\Pi_{\rm max}$ for stars of various mass.
Figure~6 depicts the driving regions for three different model sequences
($M=0.57, 0.60$ and $0.63~M_{\odot}$) at three different 
effective temperatures (144,000~K, 100,000~K, and 78,000~K).  The
driving strength, $dk/dr$, is determined from Equation~(3), where
$k$ represents the expression inside parentheses.  The vertical axis has
not been normalized and is the same scale in all three panels.
The surface of each model is on the left, at $\tau_{\rm th}=0$.  
The vertical lines in the figure represent $\Pi_{\rm max}$, for each 
model, normalized to 1000~s in the $0.57~M_{\odot}$ model at 144,000~K.
We have made no attempt to calculate the actual value of 
$\Pi_{\rm max}$ (however, see the Appendix); the important thing is 
its changing relationship
to the depth of the driving zone with changing mass and \teff.

A couple of important trends are clear in the figure.  First, the
driving zone in models of a given mass sinks to longer $\tau_{\rm th}$,
and gets larger and stronger, with decreasing \teff.  In the absence 
of other effects, this trend would lead to ever increasing periods of 
larger and larger amplitude as \teff~decreases.  Meanwhile,
$\Pi_{\rm max}$ changes more moderately, moving toward
slightly shorter timescales with decreasing \teff~and increasing
mass.  If we make the somewhat crude assumption that the effective
driving zone consists only of those parts of the full driving
zone with $\tau_{\rm th} < \Pi_{\rm max}$, then a picture of the
red edge emerges.  At \teff~=144,000~K, the driving
zone is relatively unaffected in all three model sequences.  As
\teff~decreases, and the driving zone sinks to longer $\tau_{\rm th}$,
the $\Pi_{\rm max}$-imposed limit encroaches on the driving
zone more and more for every mass.  Thus the longer periods, while
driven, are eliminated, moving the locus of power to shorter
periods than would be seen if $\Pi_{\rm max}$ was not a factor.
Eventually, {\it all} the driven periods are longer than the
maximum period at which a star can respond, and pulsation 
ceases altogether.  This is the red edge.

Pulsations will not shut down at the same \teff~in stars of every mass.  
High mass models retain more of their effective driving zones than low 
mass models at a given \teff.  This occurs because at a given \teff, the 
top of the driving zone moves toward the surface (to smaller 
$\tau_{\rm th}$) with increasing mass.  This effect of decreasing 
$\tau_{\rm th}$ at the top of the driving zone outstrips the trend to 
lower $\Pi_{\rm max}$ with increasing mass.  The result is that, at 
78,000~K, the driving zone in the $0.57~M_{\odot}$ model (upper panel of 
Figure~6) has moved to timescales entirely above the $\Pi_{\rm max}$ 
limit, while the $0.63~M_{\odot}$ model (appropriate 
for PG~0122: see the bottom panel in Figure~6) still produces significant 
driving at thermal timescales below $\Pi_{\rm max}$.  

Recall from Table~1 and Figure~2 that GW~Vir stars of lower \teff~have
higher mass.  Why?  We can now give an answer: at low \teff, the low mass 
stars have stopped pulsating because they only drive periods longer than those 
at which they can respond to pulsation!  Higher mass stars have shallower
ionization zones (at a given \teff) that still drive periods shorter than the maximum 
allowed $g$-mode period, even at low \teff.  This causes the red edge to move to 
higher mass with decreasing luminosity and \teff, {\it the same trend followed by lines
of constant period spacing.}  The interplay between driving zone depth and
$\Pi_{\rm max}$ enforces the strikingly small range of period spacings
($\Delta\Pi \sim 21.5~$s) seen in Table~1 and Figure~2.

These calculations are not an attempt to predict the exact location 
of the red edge at a given mass.   We are only interested at this point 
in demonstrating how the position of the red edge is expected to vary 
with mass, given an ionization-based driving zone with an upper limit 
placed on it by $\Pi_{\rm max}$.  In the particular models 
shown in Figure~6, the top of the driving zone corresponds to the 
ionization temperature of OVI, but the behavior of the red edge should be 
the same no matter what species causes the driving.  Its absolute location, 
though, would probably be different given driving by different species.  
In order to use the location of the red edge for stars of different mass 
to identify the exact species that accomplishes driving, we would need 
to calculate $\Pi_{\rm max}$ precisely for all the models in Figure~6 
and better understand exactly how the value of $\Pi_{\rm max}$ affects 
the amplitudes of modes nearby in period.  Such a calculation is beyond 
the scope of this paper, and we leave it to future investigations to 
attempt one.

Alternatively, the discovery of more GW~Vir stars would help us better
understand these processes by defining the red edge with greater 
observational precision.  The simplest test of our theories would
be to find low-\teff~GW~Vir pulsators of low mass.  If they exist,
then $\Pi_{\rm max}$ is probably not a factor in determining their
periods, and variation should be sought in their lightcurves at
longer periods than in the GW~Vir stars found so far; Figure~6
suggests their dominant periods should be in the thousands of seconds.
It is possible that these stars could be quite numerous (and {\it should}
be quite numerous if they exist at all, given their slower rate of
evolution) and still escape detection, since standard time-series 
aperture photometry is not generally effective at these timescales.  
Most current CCD photometers are quite capable of searching for 
such variability, however.  Based on these results, we 
encourage future studies to determine whether or not low-mass, 
low-\teff~GW~Vir pulsators do indeed exist.

\section{Summary and Conclusions}

Our purpose is to understand how and why PWD stars pulsate, so that astronomers
can confidently apply knowledge of PWD structure---gained via asteroseismology---to
study how white dwarfs form and evolve and better understand the physics that
governs these processes.  We have pursued PWD structure via the
pulsation periods and the additional constraints they provide for our models.
A surprising similarity emerged among them:
their patterns of average period spacing span a very small range from
21--23~s.  Including the PNNV star NGC~1501, this uniformity extends
over three orders of magnitude in PWD luminosity.  Since the average period 
spacing increases with decreasing luminosity---and decreases with
increasing mass---this result implies a trend toward significantly higher
mass with decreasing luminosity through the instability strip.   This trend
has several important implications for our understanding of PWD
pulsations.  

The instability is severely sloped toward lower \teff~with stellar mass increasing 
down the instability strip.  To understand this sloped instability strip, we 
needed information inherent in the other fundamental observed trend: the 
dominant period in PWD pulsators decreases with decreasing luminosity.
We found that the observed dominant period is correlated with the theoretical
maximum $g$-mode
period, $\Pi_{\rm max}$.  If $\Pi_{\rm max}$ is a factor in determining
the range of periods observed in pulsating PWD stars, then
it should play a role in determining when pulsation ceases, because the
driving zone tends to drive longer and longer periods in models of given
mass as the models evolve to lower \teff.  At low enough temperatures,
the driving zone is only capable of driving periods longer than
$\Pi_{\rm max}$, and pulsations should then cease.  Since the top of the
driving zone moves to shorter timescales with increasing mass, and does
so faster than $\Pi_{\rm max}$ {\it decreases} with increasing mass, higher mass
models should pulsate at lower \teff~than lower mass models.  This
behavior is compatible with the observed slope of the red edge in the
log~$g$--log~\teff~plane.

This mechanism does not account for the lack of observed high-mass pulsators
at high~\teff, however.  Theoretical luminosity functions for the PWD stars
indicate the observed blue edge is probably significantly affected
by selection effects due to rapid evolution at high \teff.   Since higher mass
models are both less numerous and less luminous at a given \teff~than
models of lower mass, we are more likely to detect low-mass than high-mass
PWDs at a given temperature.  This will cause stars to ``bunch
up'' against the red edge in the observed instability strip, and the apparent blue
edge will thus ``artificially'' resemble the shape of the red edge---no matter
what the true shape might be.  This selection effect, in isolation, would imply
that low-mass, low \teff~GW~Vir pulsators should be most numerous of all.  That we
in fact find {\it none} strengthens our contention that the mass-dependence
of the red edge is a real effect---we find no low-mass, low-\teff~GW~Vir
stars because they don't exist.

The boundaries of the PWD instability strip derived from
our pulsation studies are far smaller than those based on spectroscopic
measurements alone.  Though the actual range of log~$g$ and \teff~spanned
by the known pulsators is no smaller than before, the newly discovered
mass-luminosity relationship implies that the width of the strip
in \teff~{\it at a given log~$g$} is quite small, and vice versa.
We can therefore no longer say for certain that {\it any} non-pulsators
occupy this newly diminished instability strip, since the uncertainties in
log~$g$ and \teff~as determined from spectroscopy are larger now than
the observed width of the strip itself at a given log~$g$ or \teff.

This highlights the importance of finding additional pulsators with
which to further refine our knowledge of the instability strip boundaries.
In particular, we still have no observations with which to
constrain theories of the blue edge, since there is no reason that the 
{\it instrinsic} blue edge lies near the {\it observed}
blue edge at any effective temperature.

If the trend we have discovered continues down to effective temperatures below the
coolest known pulsating PWDs, then high-mass ($\sim 1~M_{\odot}$ or greater) 
white dwarfs
might pulsate at temperatures as low as 50,000--60,000~K
(and log~$g \sim 8$).  Their dominant periods (again, assuming they follow the
trends outlined in \S~3 and \S~5)
would be shorter than any known PWDs, perhaps as low as 200-300~s.  Such
stars would not be PWDs at all but rather white dwarfs proper.  Their
discovery would complete a ``chain'' of variable stars from PNN stars to
``naked'' PWDs to hot white dwarfs, and as such they would represent
a incalculable boon to astronomers who study the late stages of stellar
evolution.

On the theory side, we need to understand how PWD stars react
to driving near the maximum $g$-mode period.  Studies should be undertaken to
determine the actual maximum period in PWD models.  PWD evolution
sequences should be constructed that contain all the elements thought to exist in
PWD stars.   We have observational information to test all these
calculations, and with the observational program proposed above, we will gain
more.  There is much to do.

\acknowledgments

The author expresses deepest gratitude to Steve Kawaler and Chris Clemens,
who together taught him the crafts of astronomy: theory, observation, and 
communication.  Their sage council and unblinking criticism greatly enhanced 
the quality of this work.  

I am also indebted to Paul Bradley, whose thoughtful insights
substantially improved both the content and presentation of this paper.

\appendix

\section{The Maximum Period}

A variable pre-white dwarf star is a resonant cavity for non-radial
$g$-mode oscillations.  This spherically symmetric cavity is bounded by 
the stellar center (or the outer edge of the degenerate core in ZZ~Ceti
and DBV white dwarfs) and surface.  At sufficiently long periods, however,
the surface layers no longer reflect internal waves.  The pulsation
energy then leaks out through the surface, damping the pulsation.  This
idea was first applied to white dwarf pulsations by Hansen, Winget \& Kawaler 
(1985), who attempted to calculate the approximate critical frequencies to 
explain both the red edge and maximum observed periods in ZZ~Ceti stars.
Assuming an Eddington gray atmosphere, they derive the following expression 
for the dimensionless critical $g$-mode frequency:
\begin{equation}
\omega_{g}^{2} \approx \frac{\ell(\ell+1)}{V_{g}}
\end{equation}
where
\begin{equation}
V_{g} = \frac{3 g \mu R}{5N_{a}kT_{\rm eff}}.
\end{equation}
Here $g$ and $R$ represent the photospheric surface gravity 
and radius, and $N_{a}$, $k$, and $\mu$ have their usual meaning.

The dimensionless frequency, $\omega$, is related to the 
pulsation frequency, $\sigma$, according to
\begin{equation}
\omega^{2} = \frac{\sigma^{2}R}{g}.
\end{equation}
The pulsation period is $\Pi = 2\pi / \sigma$, so combining 
Equations~(A1) through (A3) we arrive at the maximum $g$-mode 
period:
\begin{equation}
\Pi_{\rm max} \approx  940s~\left(\frac{\mu}{\ell(\ell+1)}\right)^{0.5} 
\left(\frac{R}{0.02R_{\odot}}\right) 
\left(\frac{T_{\rm eff}}{10^{5}{\rm K}}\right)^{-0.5},
\end{equation}
or, using the relation $R^{2} = L / 4\pi \sigma T_{\rm eff}^{4}$,
\begin{equation}
\Pi_{\rm max} \approx 940s~\left(\frac{\mu}{\ell(\ell+1)}\right)^{0.5} 
\left(\frac{L}{35L_{\odot}}\right)^{0.5} 
\left(\frac{T_{\rm eff}}{10^{5}{\rm K}}\right)^{-2.5}.
\end{equation}

For PG~1159, with $L = 200~L_{\odot}$ and $T_{\rm eff} = 140,000$~K, 
Equation~(A5) predicts $\Pi_{\rm max} = 850$~s for $\ell=1$ modes 
and 492~s for $\ell=2$ modes.\footnote {Assuming a mix of C:O:He = 
0.4:0.3:0.3 by mass, implying $\mu=1.59$.}  PG~0122, with 
$L = 5.6~L_{\odot}$ and $T_{\rm eff} = 80,000$~K, would have 
$\Pi_{\rm max} = 580$~s for $\ell=1$ and 340~s for $\ell=2$. Because
of the simplicity of the gray atmosphere assumption, these numbers 
are more useful in comparison to each other than as quantitative 
diagnostics of the maximum period. For instance, Hansen, Winget 
\& Kawaler (1985) find that the values of $\Pi_{\rm max}$ derived 
from this analysis are ``within a factor of 2'' of those based on 
more rigorous calculations.  We are more interested here in the run
of $\Pi_{\rm max}$ with respect to global stellar quantities such 
as $L$ and $ T_{\rm eff}$.

If $\Pi_{\rm max}$ determines the long-period cutoff in GW~Vir 
pulsators, then the longest period $\ell=1$ modes should be 
(approximately) a factor of 1.73 times longer than the longest
period $\ell=2$ modes.  PG~1159 is the only GW~Vir star with 
positively identified $\ell=2$ modes.  In the period list of 
Winget et al.~(1991), the longest period (positively identified) 
$\ell=2$ mode has a period of 425~s, while the longest period 
$\ell=1$ mode has a period of 840~s.  The ratio of these two periods 
is 1.98, close to the predicted ratio.  The periods themselves are 
also surprisingly close to the calculated values.  

The predicted ratios between the longest period modes also hold for 
intercomparison of different stars.  The longest period mode so far 
identified for PG~0122 is 611~s, 1.37 times smaller than the longest 
period in PG~1159, again very close to the predicted ratio of 1.43 
from Equations~(A4) and (A5).  The agreement between $\Pi_{\rm max}$ 
from our rough calculations and the observed maximum periods in 
GW~Vir stars is impressive enough to warrant further study.  In 
particular, more rigorous theoretical calculations of $\Pi_{\rm max}$ 
should be undertaken to corroborate or refute these results. 

\clearpage

\begin{table}
\begin{center}
\vspace*{-0.5in}
\caption{Summary of Variable PWD Parameters}
\vspace*{0.2in}
\begin{tabular}{lccccc}

\tableline
\tableline

Name & ~~log~$g$~~ & ~~\teff~~ & ~~$log(L/L_{\odot})$~~ &~~$M/M_{\odot}$~~ & $\Delta\Pi$ \\
\tableline
NGC~1501 & $\sim 6.0$ & $80\pm10~$kK & $3.3\pm0.3$ & $0.55\pm0.03$ & $\sim~22.3~$s \\
 & &  &  &  \\
PG~1159 & $7.22\pm0.06$ & $14\pm10~$kK & $2.3\pm0.2$ & $0.57\pm0.01$ & $21.5\pm0.1~$s \\
 & &  &  &  \\
PG~2131 & $7.67\pm0.12$ & $95\pm10~$kK & $1.0\pm0.2$ & $0.61\pm0.02$ & $21.6\pm0.4~$s \\
 & &  &  &  \\
PG~0122 & $7.97\pm0.15$ & $80\pm10~$kK & $0.7\pm0.2$ & $0.68\pm0.04$ & $21.1\pm0.4~$s \\
\tableline
\end{tabular}
\end{center}
\end{table}

\clearpage

\begin{figure}[h]
\vbox to5.6in{\rule{0pt}{5.6in}}
\includegraphics{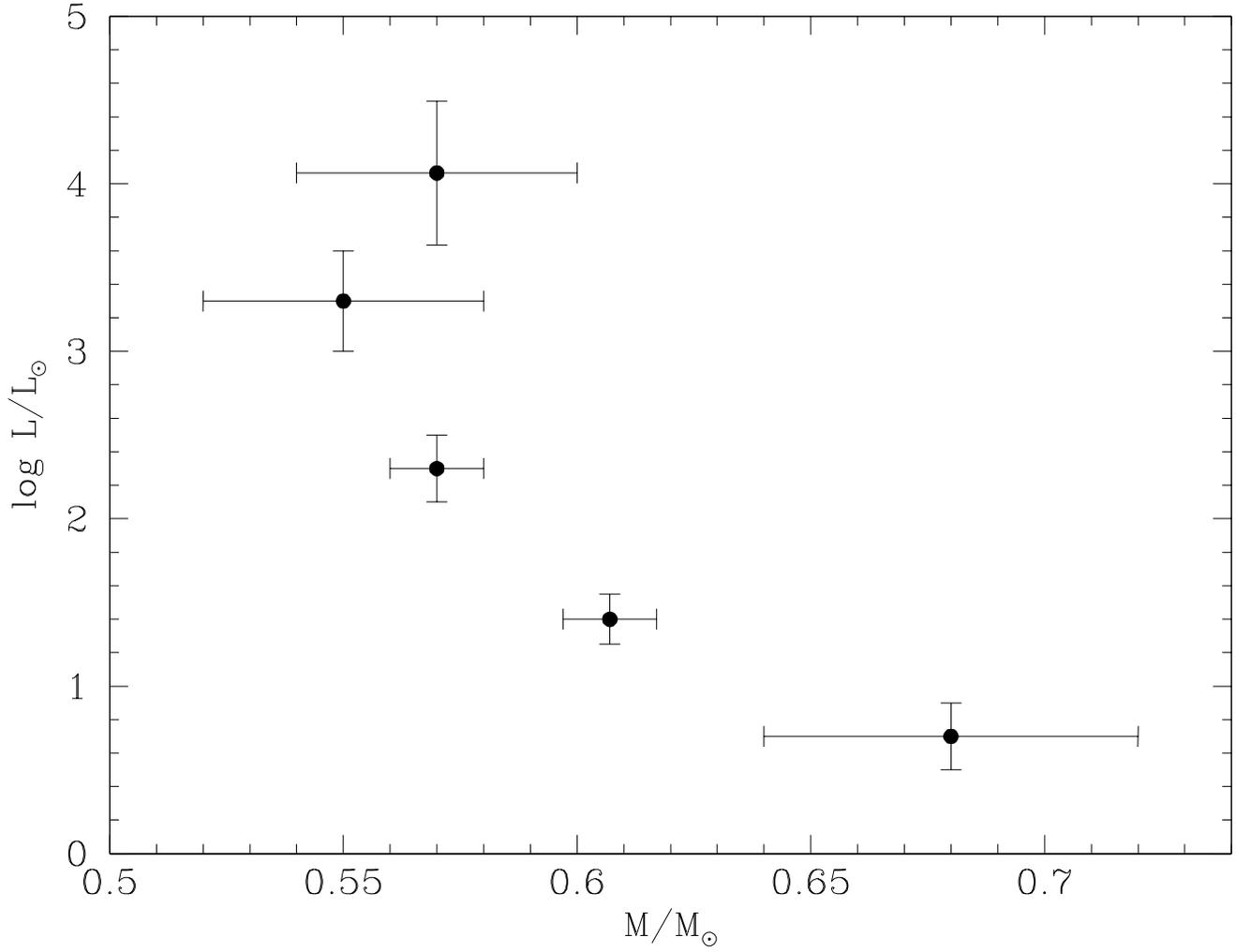}
\caption{Luminosity versus mass determined from pulsation data for the
GW~Vir stars PG~0122, PG~2131 and PG~1159, and for two ``evolved'' PNNV stars,
NGC~1501 and RXJ~2117.}
\end{figure}

\clearpage

\begin{figure}
\vbox to5.6in{\rule{0pt}{5.6in}}
\includegraphics{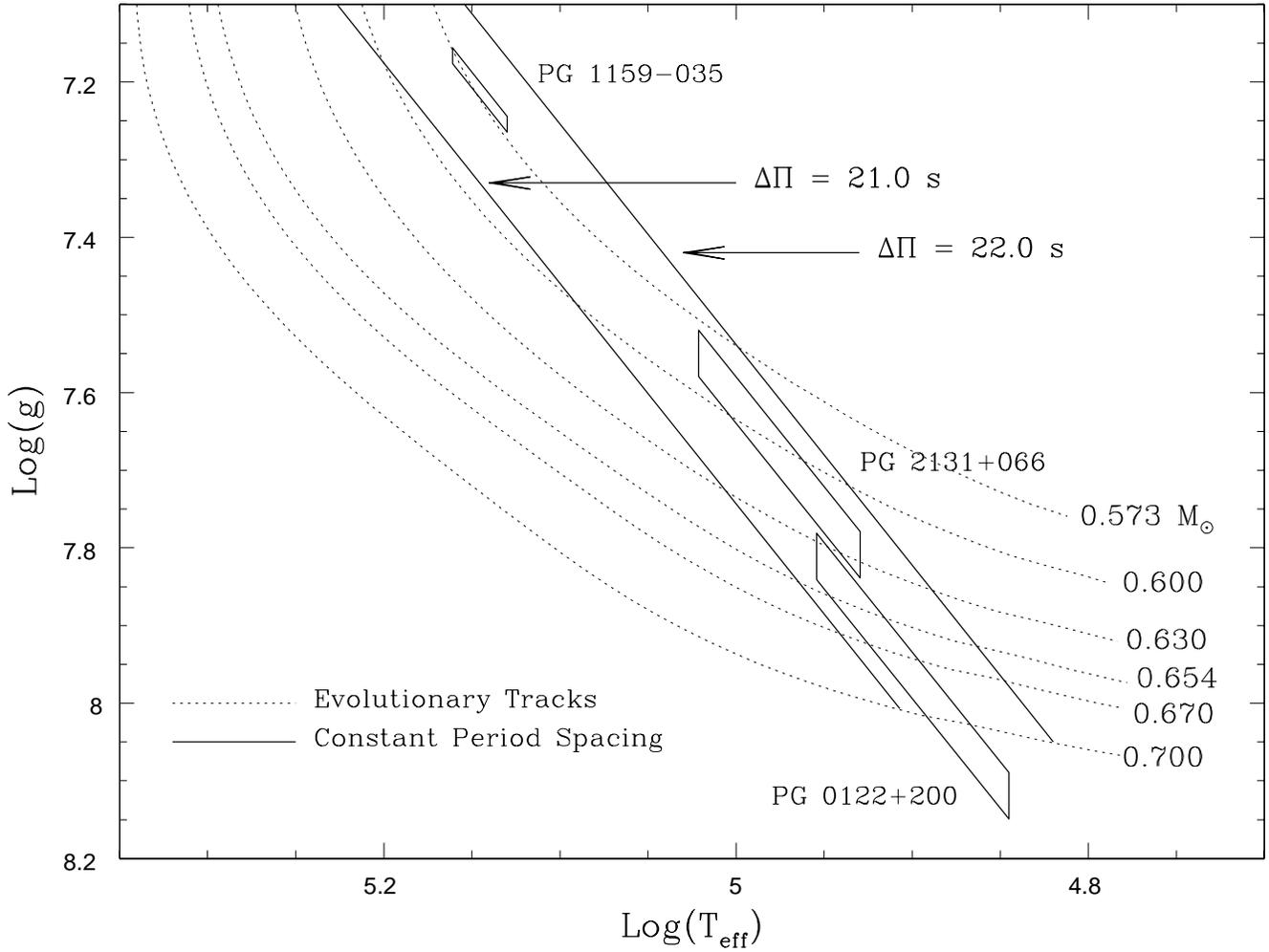}
\caption{Portion of the log~$g$--log~\teff~plane showing PWD
evolutionary tracks based on ISUEVO.  The error boxes for each
star derive from period spacing combined with spectroscopically 
determined \teff.}
\end{figure}

\begin{figure}
\vbox to7.0in{\rule{0pt}{7.0in}}
\includegraphics{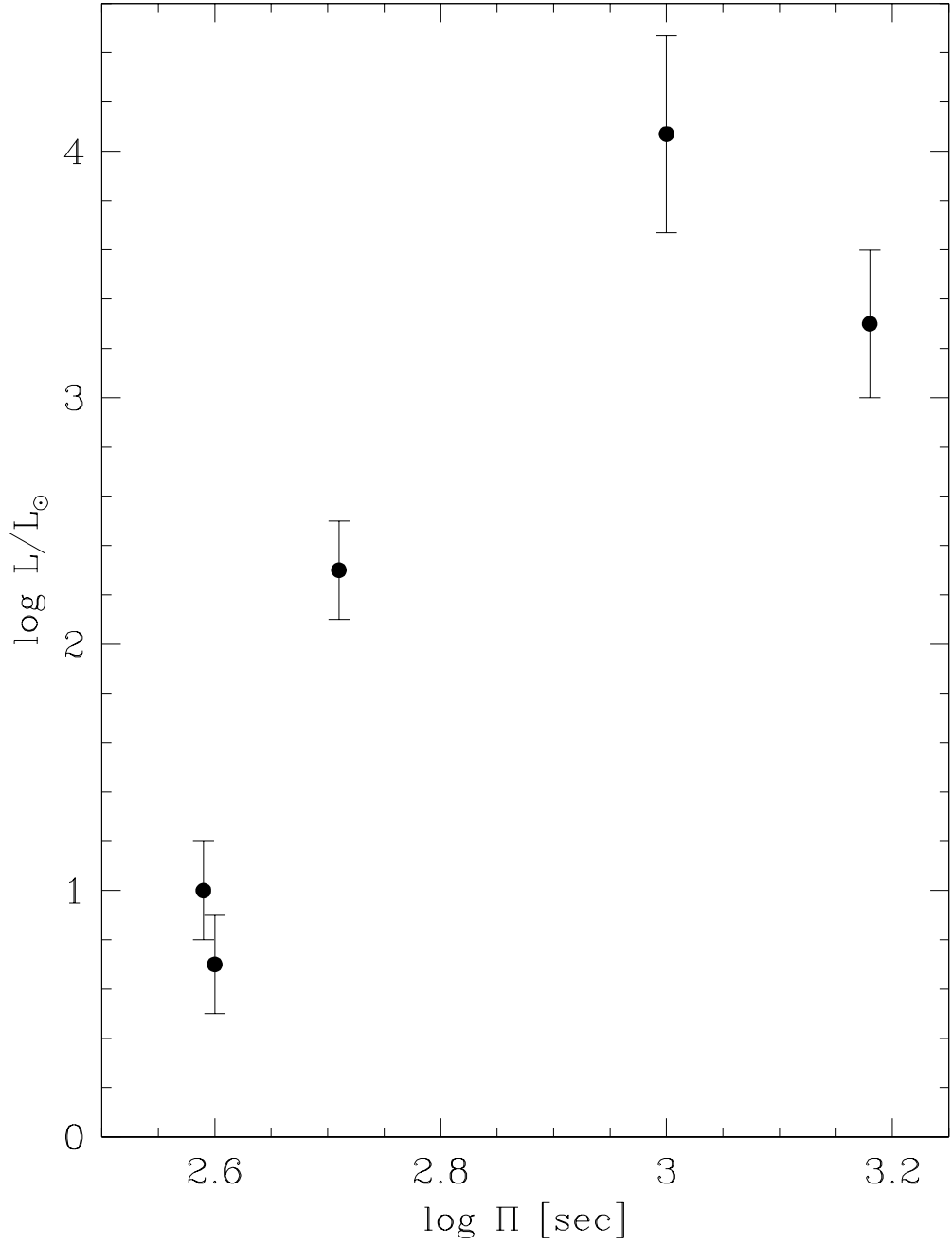}
\caption{Luminosity (determined using the pulsation data) versus
dominant period for the same PWD stars as in Figure~1.}
\end{figure}

\begin{figure}
\vbox to5.6in{\rule{0pt}{5.6in}}
\includegraphics{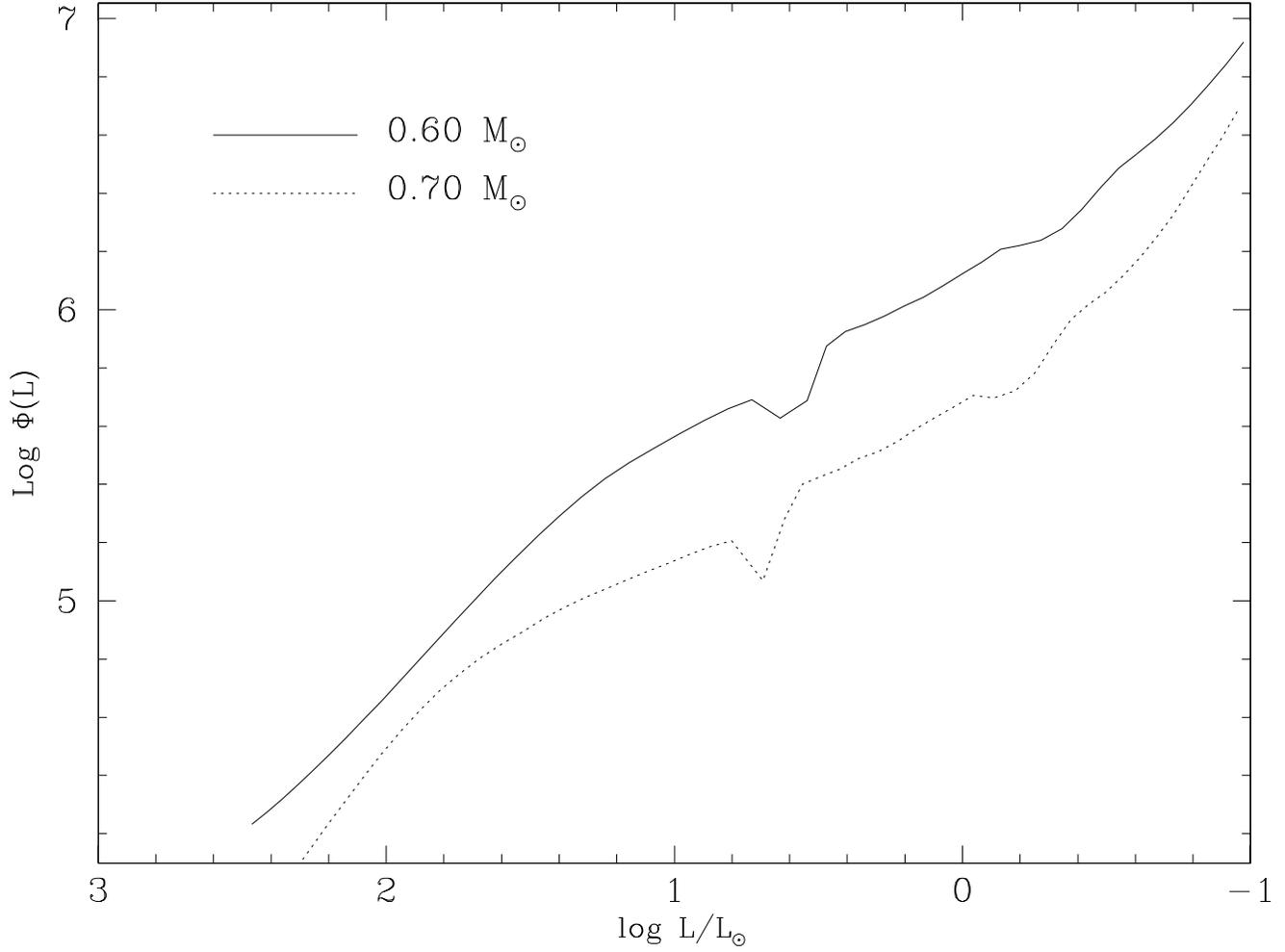}
\caption{Theoretical luminosity functions, log~$\Phi(L)$, for 
two different single-mass populations of PWD stars.}
\end{figure}

\begin{figure}
\vbox to5.6in{\rule{0pt}{5.6in}}
\includegraphics{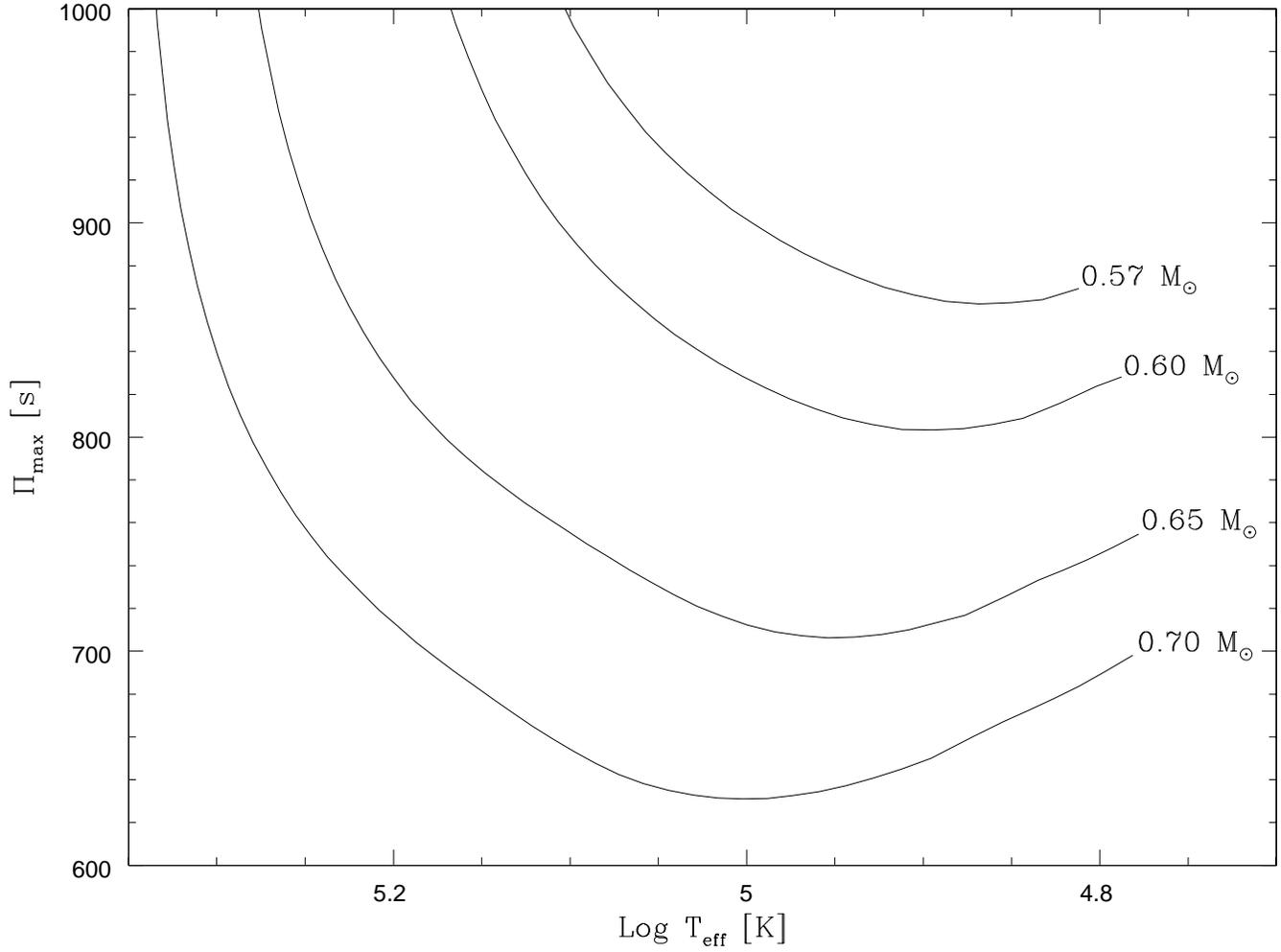}
\caption{Maximum theoretical $g$-mode period, $\Pi_{\rm max}$, 
{\it independent of the driving period}, versus \teff~for
four PWD evolution sequences of different mass.}
\end{figure}
\clearpage

\begin{figure}
\vbox to7.0in{\rule{0pt}{7.0in}}
\includegraphics{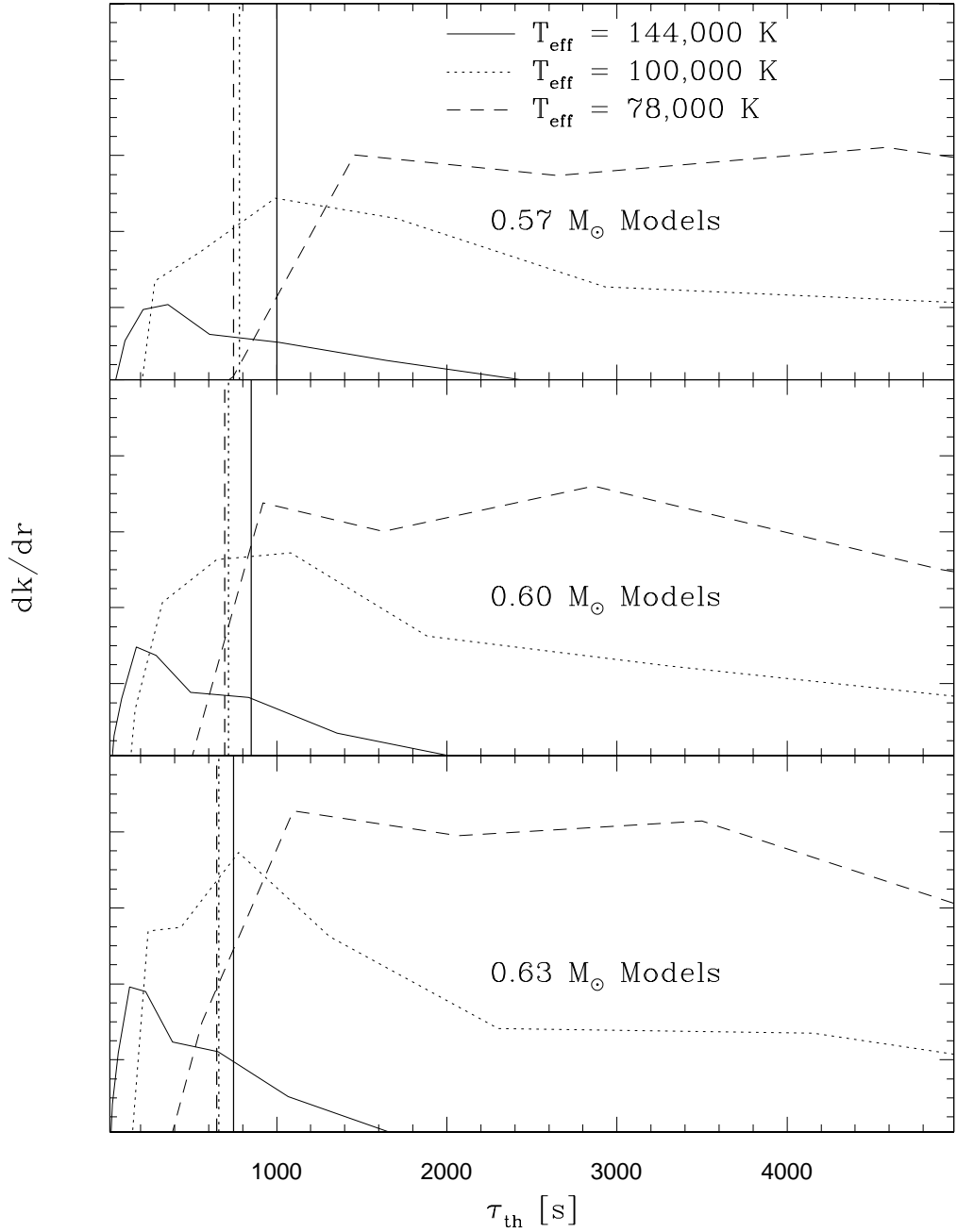}
\caption{Driving strength, $dk/dr > 0$, for three different model 
sequences at three different effective temperatures. Vertical 
lines represent the maximum $g$-mode period at which the star can 
respond to driving.}
\end{figure}

\end{document}